\documentclass[a4paper,11pt]{article}
\usepackage{jcappub} % for details on the use of the package, please
\usepackage{graphicx,hyperref}
\usepackage{amssymb}
\usepackage{epstopdf,color}
\usepackage{color}
\newcommand{\GeV}{\text{ GeV}}
\newcommand{\MeV}{\text{ MeV}}
\newcommand{\keV}{\text{ keV}}
\newcommand{\meV}{\text{ meV}}

\usepackage{comment}
\newcommand{\req}[1]{Eq.\,(\ref{#1})}

\title{\boldmath Dynamical Emergence of the Universe into the False Vacuum}
\author{Johann Rafelski}
\author{and Jeremiah Birrell}
\affiliation{Department of Physics, University of Arizona,  Tucson,  Arizona, 85721, USA}
\date{September 12, 2015}

\abstract{We study how the hot Universe evolves and acquires the prevailing vacuum state, demonstrating that in specific conditions which are believed to apply,  the Universe becomes frozen into the state  with the smallest value of Higgs vacuum field $v=\langle h\rangle$, even if this is not the state of lowest energy. This supports the false vacuum  dark energy $\Lambda$-model. Under several likely hypotheses we determine the  temperature in the evolution of the Universe at which two vacuua $v_1, v_2$ can swap between being  true and false. We evaluate the dynamical surface pressure on domain walls between low and high mass vaccua due to the presence of matter and show that the  low mass state remains the preferred vacuum of the Universe.}

\begin{document}
\toccontinuoustrue
\maketitle
printed \today
\flushbottom

\section{Introduction}
 This work presents relatively simple arguments for why the cosmological evolution selects the vacuum with smallest  Higgs  VEV $v=\langle h\rangle$ which, in general, could be and likely is the `false' vacuum.  Our argument relies on the Standard Model (SM)  {\em minimal coupling: $m\to gh$}, or similar generalizations in   `beyond' SM (BSM), so that the vacuum with the smallest Higgs VEV also has the smallest particle masses.  In anticipation of the model with multiple vacuua, we call the vacuum state with lowest free energy at temperature $T$ \lq the true vacuum\rq\ and all others \lq the false vacuua\rq.  Note that this is a temperature dependent statement: we live today  in the false vacuum which as we will show was once the true vacuum. 

In the presence of  pairs of particles and antiparticles at  high temperature the vacuum state with smallest $v $ is energetically preferred, even if it has a large vacuum energy.  This is so because smaller $v$ implies smaller particle masses and hence less energy, and free energy, in the particle distributions.  By the time the Universe cools sufficiently for the larger vacuum energy to dominate the smaller particle free energies, the probability of swap to the large mass  true vacuum  is vanishingly small in general.

Therefore, the Higgs minimum with the lowest value of the Higgs field $v$, and thus {\em  not necessarily} the lowest value of the effective potential $W(v)=\langle V(h)\rangle$, emerges as the prevalent vacuum in our Universe. The   difference, $\rho_{\Lambda}=\Delta W$, between the prevalent vacuum state  today and the true minimum is a natural candidate to explain the observed dark energy density,
\begin{equation}\label{de_density}
\rho_{\Lambda}=25.6\meV^4.
\end{equation}
For further theoretical discussion of dark energy  as a vacuum property see Ref.\cite{Padmanabhan:2002ji}, see also Refs.\cite{Peebles:2002gy,Weinberg:1988cp}. The  value of dark energy we quote in \req{de_density} is derived from  Ref.\cite{pdg2014}. The essential observational requirement for the $\Lambda$-vacuum model of dark energy is the determination of time independence and space homogeneity of $\rho_{\Lambda}$. 

Section~\ref{heavyM} establishes the required elements of the standard model (SM) and shows in detail  the effect of matter,  for the most part highly abundant pairs of particles present in the hot Universe. The insights  we present in section~\ref{heavyM} are already present in the vast literature exploring the EW phase transitions, a different topic, but with similar technical context. Specifically:\\
\indent i) Phase transitions involving change in the vacuum state are a prime preoccupation in particle cosmology, and requires study of  the nature of the phase transition between the stages of hot SM matter, as it can be the source of baryogenesis, see for example Ref.\cite{Kuzmin:1985mm,Shaposhnikov:1987tw,Anderson:1991zb}.\\ 
\indent ii) More recently another related topic, the instability of the vacuum state has  captured much attention: the study of  vacuum metastability is seen in the evaluation of the connection of potentially metastable multi-minimum SM to cosmology in Refs.\cite{Espinosa:2007qp,Ellis:2009tp,Hook:2014uia,Kearney:2015vba}. As the opening phrase in the summary of the very recent Ref.\cite{Espinosa:2015qea}  succinctly observes:  \lq\lq Assuming that the SM holds up to large energies, we studied under which conditions the cosmological evolution does not disrupt the electroweak vacuum, in spite of the presence of an instability of the SM effective Higgs potential $V (h)$ at field values $h > h_\mathrm{max}$\rq\rq. This comprehensive study of the connection of the Higgs vacuum to cosmology  does not address the mechanisms that govern the dynamical selection of the false vacuum.
 
 We characterize the need to study how the Universe evolves into and remains captured in the false vacuum in subsection \ref{General}. In subsection~\ref{leptoswap}  we  address  this swap between false and true vacuum quantitatively including pairs of particles and antiparticles at finite temperature and for electrons at finite chemical potential. This is a static equilibrium argument. In subsection~\ref{surfaceP} we add the dynamics of particle exchange between large space domains of true and false vacuum. This provides  a  dynamical pressure on domain walls separating regions of different vaccua that, in general, differs from the bulk pressure differential between the two vacuua.  We investigate this surface effect  and show that in our context it does not alter the conclusions.

%%%%%%%%%%%%%%%%%%%%%%%%%%%%%%%%%%%%%%%%%%%%%%%%%%%%%%%%%%%%%%
\section{ Effective Potential in the Hot Universe}\label{heavyM}
%%%%%%%%%%%%%%%%%%%%%%%%%%%%%%%%%%%%%%%%%%%%%%%%%%%%%%%%%%%%%%%%%%%%%%%%%%%%%%%%%%%%%%%%%%%%
\subsection{Higgs Effective Potential $W(v)$}
A double nearly degenerate vacuum structure was anticipated in the literature but received for many years only scant recognition. In 1995 Froggatt and Nielsen predicted the correct value of both the  top quark mass and the Higgs mass by arguing that the  Higgs potential should have  two degenerate vacuum states~\cite{Froggatt:1995rt}.  A somewhat different line of thought led Shaposhnikov and Wetterich to the rediscovery of this set of masses in 2010~\cite{Shaposhnikov:2009pv}. The mechanism of  Froggatt and Nielsen requires that the parameters which are known to govern the SM, in particular the value of the masses of both the top quark and the Higgs, indicate that vacuum fluctuations (running of SM parameters) lead to a modification of the SM effective  vacuum  potential $W(v)$ to the degree that the rising $h^4$ behavior of the SM Higgs potential is compensated, leading to an instability, that is a fall-off for large $v$~\cite{Masina:2015ixa,Blum:2015rpa,Branchina:2014rva,Branchina:2014usa,Zoller:2014cka,Kobakhidze:2014xda}. Such vacuum meta-instability, also in view of Froggatt and Nielsen, can  be cured~\cite{Branchina:2014rva,Kobakhidze:2014xda,EliasMiro:2011aa}, resulting in a landscape in $W(v)$ with   two minima, albeit with the 2nd one at an extreme Plankian scale value of $v=\langle h\rangle$ -- this being a consequence of the  absence of any  scale other than $v=v_H =246\GeV$  and the Plank mass $M_P\gg v_H$. 
 
In our investigation, we  use a simple model potential  consisting of vacuum states at $\pm v_1=v_H$ and $\pm v_2$,
\begin{equation}\label{eff_pot_model}
W(v)=\frac{m_H^2}{8v_1^2}(v^2-v_1^2)^2\frac{(v^2-v_2^2)^2}{(v_1^2-v_2^2)^2}.
\end{equation}
The coefficient is chosen so that at $\pm v_1=v_H$, the Higgs mass takes its measured value.  

Another  way to understand the proposed form of the potential \req{eff_pot_model} is to note that when  $v_2\gg v_1$ and  $|v|\ll v_2$ this expression reduces to the  usual Higgs SM. The Higgs part of the SM bi-quartic Higgs field Lagrangian is written in the form 
\begin{align}\label{effective_lagrangian}
\mathcal{L}_H=&\frac{1}{2}\partial_\mu h\partial^\mu h-V(h),\hspace{2mm}V(h)=\frac{m_H^2}{8v_H^2}(h^2-v_H^2)^2,\\
 v_H=&246.2\GeV,\hspace{2mm} m_H=125.1\GeV.
\end{align}
with the zero temperature effective potential 
\begin{equation}\label{Wvpot}
W(v)\equiv \langle V \rangle,\quad v\equiv \langle h\rangle\;.
\end{equation}

%%%%%%%%%%%%%%%%%%%%%%%%%%%%%%%%%%%%%%%%%%%%%%%%%%%%%%%%%%%%%%%%%%%%%%%%%%%%%%%%%%%%%%%%%%%%
\subsection{Including Hot Matter in the SM Effective Potential}
At finite temperature  each particle species contributes a free energy density term $F=(-T\ln Z)/V$ to the Higgs effective potential, $W(v)$.   The free  energy density terms for equilibrium fermion and boson gases are seen in many textbooks
\begin{align}\label{F_no_shift}      
F_{\pm}(m,T)=&\mp\frac{d T}{2\pi^2}\int_0^\infty\!\!\!\! \ln\left(1\pm e^{-(E-\mu)/T}\right)p^2 dp,\\
E=&\sqrt{p^2+m^2}, \hspace{2mm} m=m(v),\notag 
\end{align}
where $d$ is the degeneracy, $\mu$ is the chemical potential, the upper sign is for fermions, and the lower for bosons.   When convenient we will write $F(v,T)$ for $F(m(v),T)$.

We will define the matter-Higgs coupling of each particle species by
\begin{equation}
g=\partial_v m.
\end{equation}
 In the minimal coupling scenario $g$ is a constant, but in general it could depend on $v$. 

Integration by parts shows that, for a free Bose or Fermi gas, the free energy equals the negative of the pressure, and as indicated this is consistent with the thermodynamic relationship governing the total free energy ${\cal F}$ of an extensive system:
\begin{align}\label{free_energy_pressure}
F_\pm=-P; \qquad P=-\left.\frac{\partial {\cal F}}{\partial V}\right|_T\;.
\end{align}

The finite temperature vacuum states are the critical points, $v_k(T)$, of the effective potential 
\begin{equation}\label{UTpot}
U(v,T)=W(v)+\sum_j F_j(v,T)\;,
\end{equation}
obtained by solving the equation
\begin{align}\label{pot_deriv}
0=&W^\prime(v)+\sum_j\frac{d_{j}g_j m_j}{2\pi^2}\!\int_0^\infty\!\!\!\!\frac{p^2/E_j}{e^{(E_j-\mu)/T}\pm 1}\,dp.
\end{align}
%DEFINITION MOVED TO INTRO: In anticipation of the model with multiple vacuua, we call the vacuum state with smallest value of $U(v_k,T)$ the true vacuum at temperature $T$ and all others the false vacuua.  Note that this is a temperature dependent statement. 
The fact that $\partial_v F_j>0$ suggests that if the zero temperature true vacuum occurs at some large VEV then finite temperature corrections can raise its effective potential above vacuua at lower VEV's, causing, at some non-zero $T$, the swapping of true and false vacuua.

%%%%%%%%%%%%%%%%%%%%%%%%%%%%%%%%%%%%%%%%%%%%%%%%%%%%%%%%%%%%%%%%%%%%%%%%%%%%%%%%%%%%%%%%%%%%
\subsection{High Temperature Behavior: Capture in a False Vacuum}\label{false}
The Universe's choice of Higgs type vacuum state as it cools occurs at $T\approx v_H$, and it depends on the details of the potential. The barrier heights between any multiple vacuum states  that are present are of particular importance at high $T$. Generally, the potential possesses a landscape of vacuua characterized by a vacuum expectation values  (VEV's) $v_k$ having associated  particle `$j$' masses $m_{j,k}$.  The prevailing (false) vacuum has subscript `$H$' for Higgs. 

At the ${\cal O}(100\GeV)$ energy scales considered now, the small $\Delta W=W(v_1)-W(v_2)>0$ consistent with the observed dark energy density is negligible, so we did not show it in \req{eff_pot_model}.  We chose the model \req{eff_pot_model} for its simplicity and ability to characterize  the second minimum  associated with quantum instability   of the SM.  For this reason we do not worry about renormalizibility of this model. However, even the effective model should retain gauge invariance related to the $v\leftrightarrow -v$ symmetry. The form of $W(v)$ comprises as a presumed model all vacuum fluctuation effects including running of elementary quantities, and only real matter needs to be added, see \req{UTpot}.
 
\begin{figure}
\centering
\includegraphics[width=0.7\columnwidth]{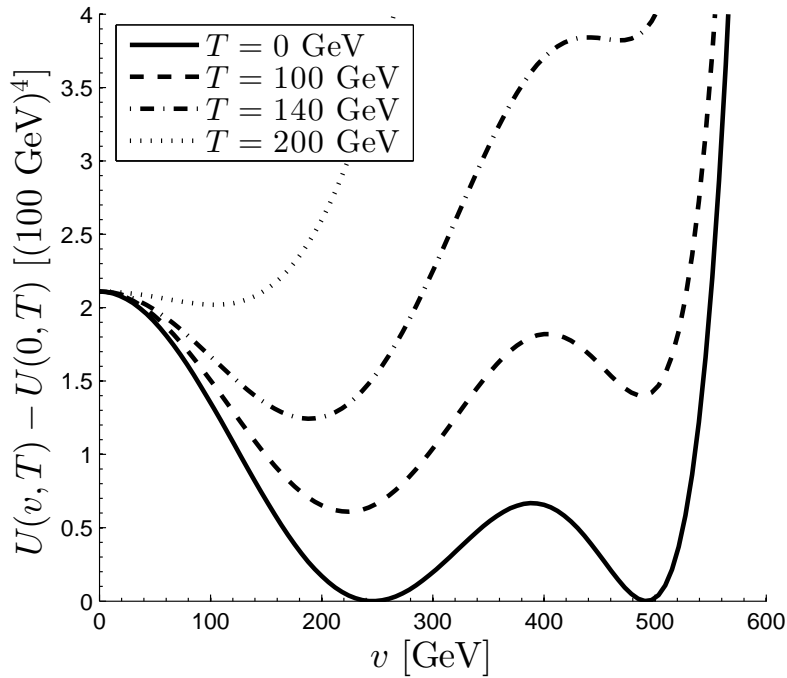}
\caption{\label{fig:Higgs_T_dep_2_vacuua} Temperature dependence of an example effective potential with vacuum states at $v_1=v_H$ and $v_2=2v_H$.}
 \end{figure}

\begin{figure}
\centering
\includegraphics[width=0.7\columnwidth]{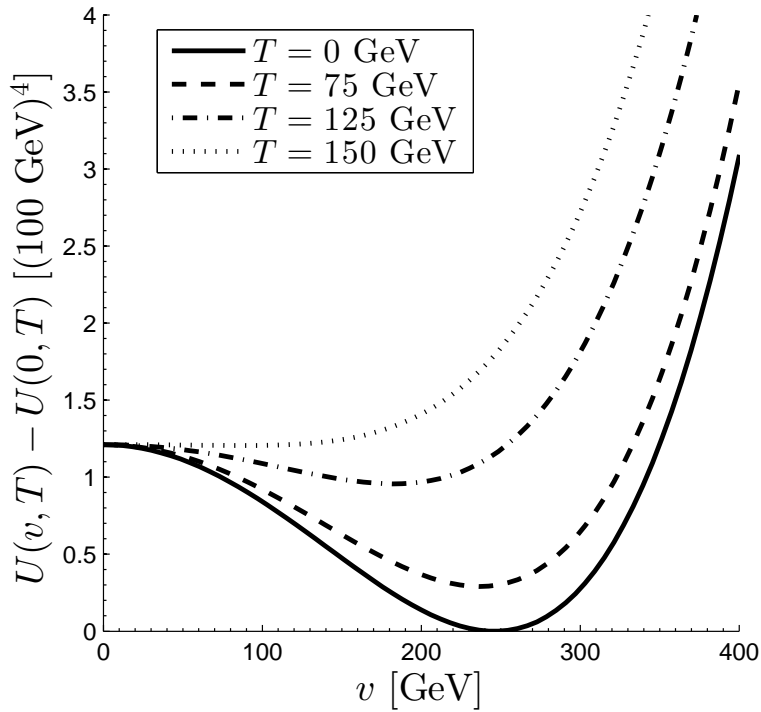}
\caption{\label{fig:Higgs_T_dep_2_vacuua_10} Temperature dependence of an example effective potential with vacuum states at $v_1=v_H$ and $v_2=10v_H$.}
 \end{figure}

We illustrate the temperature dependence for a gas of top quarks and $W$ and $Z$ bosons in  Figs.  \ref{fig:Higgs_T_dep_2_vacuua} and  \ref{fig:Higgs_T_dep_2_vacuua_10}.  All lighter particles contribute insignificantly to the shape  of the  effective potential \req{UTpot}, \req{eff_pot_model}. In Fig. \ref{fig:Higgs_T_dep_2_vacuua} we use  $v_2=2v_1$ and in Fig. \ref{fig:Higgs_T_dep_2_vacuua_10}  we use $v_2=10v_1$.  For  $v_2=2v_1$ in  Fig. \ref{fig:Higgs_T_dep_2_vacuua}  we see that the vacuum at $v_1$ forms long before that at $v_2$, and so we would expect the Universe to be trapped in $v_1$, the true vacuum at high temperature, but possibly false vacuum at low temperature.  For $v_2$ larger, the effect is only magnified.
%%%%%%%%%%%%%%%%%%%%%%%%%%%%%%%%%%%%%%%
%%%%%%%%%%%%%%%%%%%%%%%%%%%%%%%%%%%%%%%

%%%%%%%%%%%%%%%%%%%%%%%%%%%%%%%%%%%%%%%
%%%%%%%%%%%%%%%%%%%%%%%%%%%%%%%%%%%%%%%

The presence of a second vacuum state at $v_2$ can affect the electroweak phase transition temperature.  In the SM Higgs case it occurs at ${\cal O}(150\GeV)$ while in the case where $v_2=2v_1$ it occurs earlier at ${\cal O}(200\GeV)$. However, this modification of transition temperature   disappears completely  for $v_2=10v_1$. Here the behavior resembles that of the SM Higgs and Fig.  \ref{fig:Higgs_T_dep_2_vacuua_10} is indistinguishable from the plot of the temperature dependence of the SM `bare' potential \req{effective_lagrangian}. In fact we do not show in Fig. \ref{fig:Higgs_T_dep_2_vacuua_10} the part of the potential near to the already distant secondary minimum.

Given our identification of  $\Delta W$ with the meV scale of dark energy, in the above discussion  we did not need to address the effect of the energy difference  between the two vacuum states $\Delta U$ (not shown in \req{eff_pot_model}.  However, it is clear that for a sufficiently large $\Delta W$ the high temperature behavior would be modified. At some point the vacuum at the smaller value $\langle h\rangle=v_1$ will no longer be stable to tunneling to the  deeper vacuum at $\langle h\rangle =v_2$, due to a large $\Delta W$ compensating for the temperature dependence of the effective potential.  The precise value of $\Delta W$ for which the Universe is no longer likely to be captured in the (low-$T$) false vacuum will depend  on the details of the effective potential. 

 In the model considered in Fig. \ref{fig:Higgs_T_dep_2_vacuua}, and more generally in models that have a similar landscape,  a value of   $\Delta W\lesssim {\cal O}((100\GeV)^4)$ allows for capture in vacuum state that is false at  zero temperature. This means that the mechanism here discussed to generate dark energy has as an upper bound a value of the dark energy density  that exceeds the observed value by 56 orders of magnitude (4th power of 100 GeV vs meV).  Thus, depending on the landscape, this model has the potential to reduce the fine tuning that the dark energy displays,  as compared to 130 order of magnitude excess that a comparison with the zero point energy of quantum fields produces~\cite{Weinberg:1988cp}.
 
%%%%%%%%%%%%%%%%%%%%%%%%%%%%%%%%%%%%%%%
\subsection{Low Temperature Behavior}\label{lowT}
When the value of $\Delta W$ is small, the high $v$ minimum of $W$ emerges  as the true vacuum only at a relatively low temperature. Thus for a long time, as measured on a logarithmic scale,  the evolution of the Universe is in the domain that, including matter, is actually the true vacuum. In this subsection we quantify this situation. This is of importance since, in contrast to the dynamics of the Universe cooling and freezing into a particular vacuum state, the low temperature  $T\ll v_H$ dynamics are found to be largely independent of the details of the  potential.

At a finite temperature $T$, due to matter each VEV $v_k$   shifts  to $v_k(1+\delta_k)$. Linearizing the equation for $\delta_k$ we find (the differentiation is with respect to $v$)
\begin{align}
0\approx &\sum_j F^\prime_j(v_k,T)+\left[W^{\prime\prime}(v_k)+ \!\!\sum_j   F^{\prime\prime}_j(v_k)\right]v_k\delta\;.
\end{align}
Note that $V^{\prime\prime}(v_k)=m_{H,k}^2$, the square of the Higgs mass in the $k$th vacuum state.

From this we obtain the estimate
\begin{equation}
\delta={\cal O}\left(\frac{g mT^2}{v_k m_{H,k}^2 }\right)\;,
\end{equation}
where $g$ is the Higgs coupling of the dominant particle species at that temperature i.e. the heaviest particle with $m\lesssim T$. Hence, at temperatures $T^2\ll v_k m_{H,k}^2 /gm$, $\delta$ is small and such a linearization is appropriate.

Consider two vacuum states $v_1$, $v_2$ and let $\Delta U=V(v_1)-V(v_2)$. The shifted difference in the values of the effective potential at the respective vacuua are
\begin{align}
\Delta U\equiv&U(v_1(1+\delta_1),T)-U(v_2(1+\delta_2),T)\\
=&\Delta W+\sum_j F_j(v_1,T)-F_j(v_2,T)\label{deltaU_formula}\\
&+\sum_j F^\prime_j(v_1,T) v_1\delta_1- F^\prime_j(v_2,T)v_2\delta_2\notag\\
&+{\cal O}(m_{H,1}^2v_1^2\delta_1^2)+{\cal O}(m_{H,2}^2v_2^2\delta_2^2)\notag
\end{align}
where we used the fact that $V^\prime(v_i)=0$.  

Noting that 
\begin{equation}
F^\prime_j ={\cal O}(g_j(v_k) m_j( v_k)T^2)
\end{equation}
we find
\begin{equation}
m_{H,k}^2v_k^2\delta={\cal O}(g(v_k) m(v_k) v_k^2T^2)\;.
\end{equation}
This implies that both the linear and quadratic terms in $\delta$ from \req{deltaU_formula} are negligible and we have
\begin{align}\label{dU}
\Delta U=&\Delta W+\sum_j F_j(m_1,T)-F_j(m_2,T)+{\cal O}(g(v_i)m(v_i)v_i^2T^2\delta_i)\;.
\end{align}

This means that at temperatures that are small compared to the Higgs masses and VEVs $v_1, v_2$, only the difference, $\Delta W$, in the potential between the two vacuum states, along with the particle masses in the different vacuua, are significant for determining the relative heights of the effective potential in these vacuua.  The detailed structure of the potential is not important. Since $F_j$ depends on $v$ only through $m_j$, for the purposes of computing \req{dU} the precise dependence of $m_j$ on $v$ is not significant - only the masses will enter the discussion. We stress the mass dependence here since it directly controls the swap condition we discuss in the next section.

%%%%%%%%%%%%%%%%%%%%%%%%%%%%%%%%%%%%%%%%%%%%%%%%%%%%%%%%
\section{Swapping of True and False Vacuua}\label{swap}
The study of the high temperature properties and emergence of the smallest VEV $v_1$ as the `captured' true vacuum also implies that at some, lower, temperature we should expect a reversal of the situation during cosmological expansion. The formerly true vacuum becomes false, as it must be at zero temperature, and the previously false state assumes, finally, its true role. In a homogeneous Universe this reversal will hardly play an  important role. The barrier heights and widths are expected to be much greater than the difference in effective potential, and so the false vacuum is expected to be stable.

Even so, we think one cannot ignore the question -- at which temperature will this swap occur? We thus turn to discuss several different scenarios. We emphasize that this study is largely a matter of both scientific curiosity and  principle, without any eminent application or claim of Universe instability.   Therefore we present a highly abbreviated account of our exploration of vacuum swapping, pointing out the key ingredients only.

%%%%%%%%%%%%%%%%%%%%%%%%%%
\subsection{General consideration}\label{General}
The first question to consider is which family of particles present in the Universe determine  the condition of the true-false vacuum swap at fixed $v=v_1$. The lightest particle for which the  Higgs minimal coupling is presently  established with some precision is the $\tau$-lepton~\cite{AtlasTau,Choudhury:2015voa,Aad:2015gba,Khachatryan:2014jba}. The minimal coupling to $\mu^+\mu^-$ has also been confirmed, albeit at much lower precision. Therefore we use  $\tau$-leptons  to establish the upper limit on the   swap temperature of any  scenarios considered.

To see how the succession of swap temperatures arises, consider that as temperature drops heavy particles coupled thermally to photons disappear and do not influence $U(v)$. Eventually we arrive at the point where $\tau$ begins to disappear. If this is the lightest particle that couples to Higgs, once it is completely gone, the formerly true vacuum $v_1$ must certainly turn into the false vacuum, as there is no longer a particle mass dependency between the vacuua to compensate for the differences in vacuum energy. To determine the $\tau$ driven condition  we can find, by evaluating the $\tau$-pair contribution to $U(v)$ with mass $ m(v_1)$ and $m(v_2)$, where the swap occurs. Given the smallness of dark energy a small pair density of $\tau$ suffices and thus we expect $T_{s,\tau}\ll  m_\tau$.

However,   if the lighter $\mu$-leptons are also Higgs controlled, than their contribution to $U(v)$ still dominate at that temperature of the presumed $\tau$ driven swap, and the swap condition is postponed until the Universe cooled further. Again, at that muon swap condition, if electrons are Higgs controlled the swap is postponed to a lower temperature where electrons can drive the swap.  We return to discuss effect of neutrinos elsewhere \cite{Rafelski:2015lvaV1}. Photons, being massless, contribute in the same way to the free energy content in the different vacuum states and hence have no net impact.

Once the QCD confining phase forms at $T\simeq 160$ MeV,  hadrons emerge and  their mass relationship to the Higgs scale $v_H$ becomes relevant. This is subject to discussion -- some will claim that via heavy quark masses Higgs controls the scales of QCD vacuum structure. Others will argue that confinement is a low energy phenomenon unrelated to  short distance scales. For nucleons  one finds~\cite{Kronfeld:2012ym}  $m_{\rm B}\simeq m_{\rm QCD} $ and thus there could be little if any relation  to Higgs,  with light quarks having a few percent influence on baryon masses  $m_{\rm B}^2-m_{\rm QCD}^2\propto m_q^2$.  Fortunately, as we will argue, this ignorance about what controls baryon mass plays a small role in our considerations. 

For neutrinos, the relationship of mass with Higgs scale $v_H$ is another  unsettled topic:   some believe we should expect $m_\nu\propto v_H^2/M_P$, where $M_P$ is the Planck mass. If this relation is correct, swap   could be driven by neutrinos, and the details are presented in Ref.\cite{Rafelski:2015lvaV1}. However, neutrino mass is beyond the SM, and therefore  it is  possible that it  is a constant across the  vacuua we are addressing, in which case neutrinos do not contribute to the effective potential difference, and cannot drive the swap of vacuum states. Therefore we look back in history of the Universe at the next  option, the electrons and positrons,  and more generally the lepton density. 

In passing we note that Ref.\cite{Rafelski:2015lvaV1} also shows that the other free-streaming mass component in the Universe, dark matter, produces too small of a pressure to play a role in our considerations.

%%%%%%%%%%%%%%%%%%%%%%%%%%
\subsection{Swap via lepton density}\label{leptoswap} 
In the Universe we need to maintain residual electron density to neutralize the proton charge density. The asymmetry of electrons and positrons due to the net matter in the Universe  makes our evaluation slightly more complicated. The chemical potential is set by the requirement that 
\begin{equation}\label{chem_pot_eq}
(n_{e^-}-n_{e^+})/n_\gamma=\frac{\eta}{1+n_n/n_p}\equiv\tilde\eta
\end{equation}
where $\eta\approx 6.05\times 10^{-10}$ is the baryon to photon ratio~\cite{pdg2014} and the present fraction of neutrons in this number is obtained at sufficient precision from the relative abundance of $\alpha$-particles produced in  big-bang nucleosynthesis in the Universe (BBN). Relatively good agreement between observational data and theoretical results is available~\cite{Steigman:2010pa}:  $n_n \approx 0.5 n_\alpha = 0.125$, leading to $n_n/n_p=0.125/0.875\simeq 0.143$ as the present day  neutron to proton ratio.

 When the temperature is small enough, massive  particle matter densities in the Universe are dilute enough for Boltzmann statistics to apply, i.e. $(m-\mu)/T\gg 1$, and we can approximate 
\begin{equation}\label{F_low_temp}
F(m,T)\approx\!-\frac{g_d T}{2\pi^2}\!\!\int_0^\infty \!\!\!\!\!\!e^{-(E-\mu)/T}p^2dp=\!-n(m,T,\mu)T
\end{equation}
where $n(m,T,\mu)$ is the number density in the Boltzmann limit of a particle species with mass $m$, temperature $T$, and chemical potential $\mu$.  This is the well known ideal non-relativistic gas equation of state. 

We see from \req{F_low_temp} that when the temperature is too small to support  a pair density, the contribution to free energy is controlled by the prescribed conserved particle abundance; the mass of the particle does not enter.  This means that there is no contribution to $\Delta W$ from non-degenerate equilibrium particles with negligible pair density.  This is the reason why, despite their dominance of the energy density, the actual baryon abundance does not contribute to the swap condition - their pair density vanishes at much higher temperature than $e^\pm$. 

Using \req{F_low_temp}, and recalling the present day dark energy density \req{de_density}, the swap temperature is found by solving
\begin{align}\label{Tc_e_eqs}
\rho_\Lambda=&T_s[n(m_1,T_s,\mu_1)-n(m_2,T_s,\mu_2)],
\end{align}
where $n$ is the total number of $e^\pm$. Using \req{chem_pot_eq}, this can be rewritten as
\begin{align}\label{Tc_eq_e_pm}
\rho_\Lambda=&T_s[\tilde\eta n_\gamma(T_s)+2n_{e^+}(m_1,T_s,-\mu_1)-(\tilde\eta n_\gamma(T_s)+2n_{e^+}(m_2,T_s,-\mu_2))]\\
=&2T_s[n_{e^+}(m_1,T_s,-\mu_1)-n_{e^+}(m_2,T_s,-\mu_2)].\notag
\end{align}
$\mu_i>0$ are the electron chemical potentials in the two vacuua. 

The positron density, $n_{e^+}$, decays exponentially as $e^{-(m_e+\mu)/T}$ where $\mu>0$ and so the second term in \req{Tc_eq_e_pm}, with the larger electron mass due to a larger Higgs VEV, is exponentially suppressed compared to the first.  Therefore the result is largely independent of $v_2$ and the equations \req{chem_pot_eq} and \req{Tc_e_eqs} are solved by
\begin{equation}
T_s=11\keV, \hspace{2mm} \mu_1=0.22\MeV.
\end{equation}
We note that at this swap temperature, the universe is still highly homogeneous.  In addition, $T_s\ll m_e-\mu_1$ and so the use of the Boltzmann limit in \req{F_low_temp} is justified. The influence of net electron charge is noticeable; to show this we also obtained the swap condition for $e^\pm$ pair abundance only 
\begin{equation}
T_{s,\mu=0}=7.6\keV.
\end{equation}
The lower value $T_{s,\mu=0}<T_{s,\mu}$ shows that the presence of chemical potential suppresses pairs.

While it is a general conviction that all `elementary' particles of the SM receive their mass from the Higgs, in which case the above estimates close the subject, this fact is not established experimentally for the small mass particles such as the electron.  A similar analysis  applies to the heaviest $\tau$ lepton,  which we know couple to Higgs via minimal coupling. The absence of the need to look after the chemical potential related to the prevailing charge density simplifies the argument substantially.  The swap temperature is obtained by solving
\begin{align}\label{Tc_e_eqsL}
\rho_\Lambda=&T_s[n(m_1,T_s)-n(m_2,T_s)]
\end{align}
with $\mu_i=0$.  Again, the result is relatively insensitive to $m_2$  and we find
\begin{equation}
T_s=18\MeV.
\end{equation}

However, at this temperature, aside of muons,  there is a large abundance of hadrons, in particular pions, in the Universe. It is believed that the pion mass derives from chiral symmetry breaking, and hence their mass is determined by the light quark masses and a QCD vacuum factor. If  pions, and muons, and for that matter electrons as discussed above, respond to the Higgs vacuum structure, the  swap temperature is accordingly modified.  Based on the above results, when fermion pairs prevail we have  $m_f/T_s$ in the approximate range $50-100$.

%%%%%%%%%%%%%%%%%%%%%%%%%%%%%%%%%%%%%%%%%%%%%
\subsection{Dynamical Pressure on a Domain Wall}\label{surfaceP}
Surface effects at a domain wall between different vaccua lead to a dynamical, osmotic pressure on the wall that in general differs from the bulk pressure. The term osmotic pressure is used to describe a pressure differential that is created by the semi-permeability of a membrane, here the domain wall between the two vacuua.  In this case, the semi-permeability arises due to the difference in the mass of particles in one vacuum versus the other.  We have studied osmotic surface pressure and  found that this effect does not change any of our prior conclusions.

The boundary structure between  true-false vacuum domains is itself a solution involving the full nonlinear effective Higgs potential, and effects of both matter and vacuum fluctuations of all matter fields. For what follows the  relevant consideration is that for a particle near to a large domain  boundary, the boundary acts as if it were a scalar potential that modifies the particle mass, interpolating the mass value between the true and false vacuum values  $m_\mathrm{t}\gg m_\mathrm{f}$ where $m=gv$ for minimally coupled particles.

Photons can move freely across the wall, thereby equilibrating the temperature.  However,  the asymmetric nature of the wall transmissivity to particles can result in different chemical potentials on either side.  The momentum distributions in the bulk therefore take the form
\begin{align}\label{f_def}
f_i(p,\mu_i)=\frac{1}{e^{(\sqrt{m_i^2+p^2}-\mu_i)/T}\pm 1}\;,\quad i=1,2,
\end{align}
where each particle species  has mass $m_i=g_i v_i$ in the two vacuua. 

Symmetry implies that the particle energy, and the components of the four momentum tangential to the surface are conserved as a particle crosses or is reflected from the boundary.  Therefore, the particle either undergoes elastic reflection or it can pass through the wall.  If the latter occurs, kinematic considerations imply that the $4$-momenta on each side, $p_1$ and $p_2$, satisfy $p_1^2=m_1^2$ and $p_2^2=m_2^2$. This, together with conservation of energy and the tangential components of the 4-momentum yields the following relation between normal components
\begin{equation}
(p_1^\bot)^2+m_1^2=(p_2^\bot)^2+m_2^2.
\end{equation}
In particular, if $m_1<m_2$ then for a particle to cross from vacuum 1 to vacuum 2 the $\bot$-component of the four momentum must satisfy the constraint
\begin{align}\label{p_constraint}
(p_1^\bot)^2\geq m_2^2-m_1^2.
\end{align}

Consider for simplicity a domain wall lying in the $y-z$ plane with vacuum expectation values $v_1<v_2$. Assuming the domain wall is static and that scattering has negligible impact on the wall, the momentum of a particle before and after scattering is determined entirely by the above symmetry and kinematic considerations. We will formulate our general results with a momentum dependent reflection probabilities, $\rho_i(p)$, from each side of the wall but specific calculations will use a step function that simply ensures that the minimum energy threshold is met i.e.
\begin{align}\label{threshold_probabilities}
\rho_1=1_{0\leq p_x<\sqrt{m_2^2-m_1^2}},\hspace{2mm} \rho_2=0.
\end{align}
The precise nature of the reflection probabilities are beyond the scope of this work.

The net dynamical pressure on the wall is obtained by comparing the change in momentum of particles reflecting from and passing through the wall on each side
\begin{align}\label{delta P}
\Delta P=P_1-P_2=&\frac{d}{8\pi^3}\int_{p_1^x\geq 0}f_1(p_1,\mu_1)\frac{p_1^x}{E_1}(2p_1^x\rho_1(p_1)+(p_1^x-\tilde p_1^x)(1-\rho_1(p_1)))d^3p_1\\
&-\frac{d}{8\pi^3}\int_{p_2^x\leq  0}f_2(p_2,\mu_2)\frac{p_2^x}{E_2}(2p_2^x\rho_2(p_2)+(p_2^x-\tilde p_2^x)(1-\rho_2(p_2)))d^3p_2
\end{align}
where
\begin{align}
\tilde p_1^x=\sqrt{(p_1^x)^2+m_1^2-m_2^2},\hspace{2mm} \tilde p_2^x=-\sqrt{(p_2^x)^2+m_2^2-m_1^2}
\end{align}
are the final momenta of a particle traversing the wall from left to right with initial 4-momentum $p_1$ and from right to left with initial 4-momentum $p_2$ respectively.  We investigate this net pressure in  two   regimes of physical significance.

%%%%%%%%%%%%%%%%%%%%%%%%%%%%%%%%%%%%%%%%%%%%%
%\subsection{Impact on Static Phase Equilibrium Considerations}
%\subsubsection{Equal Chemical Potential}
%%%%%%%%%%%%%%%%%%%%%%%%%%%%%%%%%%%%%%%%%%%%%
When the chemical potentials on either side of the barrier are the same, $\mu_1=\mu_2=\mu$, the specific form of the threshold transmission probability \req{threshold_probabilities} results in the dynamical pressure on the domain wall, $\Delta P$, being equal to the difference in bulk pressures on either side of the wall, where the bulk pressure is given by
\begin{align}
P_{\rm bulk}=\frac{d}{6\pi^2}\int_0^\infty f(p,\mu) \frac{p^4}{E}dp.
\end{align}
In other words is the osmotic pressure effect vanishes.

This equality can be seen as follows.  We can extract the bulk pressure differential from \req{delta P} to obtain
\begin{align}
\Delta P=&\Delta P_{\rm bulk}-\frac{d}{8\pi^3}\left(\int_{p^x\geq 0}f_1(p,\mu)\frac{p^x}{\sqrt{p^2+m_1^2}}\left(p^x+\sqrt{(p^x)^2+m_1^2-m_2^2}\right)(1-\rho_1(p))d^3p\right.\\
&-\left.\int f_2(p,\mu)\frac{p^x}{\sqrt{p^2+m_2^2}}\left(p^x+\sqrt{(p^x)^2+m_2^2-m_1^2}\right)d^3p\right).
\end{align}
Changing variables $z^2=p_x^2+m_1^2-m_2^2$ in the first integral and using \req{threshold_probabilities} transforms the first integral into a form that is identical to the second, proving that $\Delta P=\Delta P_{\rm bulk}$.

Specifically, in the relativistic regime where particle chemical potentials are negligible compared to the ambient temperature, the preferred vacuum is determined by the bulk pressure differential, and hence by the free energy \req{F_no_shift}.

We emphasize that the exact cancellation is a conspiracy between the scales $m_1$ and $m_2$ that simultaneously control the particle energies on each side of the wall as well as the transition probabilities. One could envision other related models where the transition probabilities are controlled by additional scales and in these cases one would expect an osmotic effect arising from the deviation of the dynamical surface pressure from the bulk pressure differential.

The swap conditions computed in section \ref{swap} occur in the non-relativistic regime, $T\ll m_1<m_2$.  Here the dynamical surface pressure is well approximated by the bulk pressure differential for a different, and more general, reason than in the case of equal chemical potentials.  In the non-relativistic limit, almost no particles have sufficiently high momentum to satisfy the constraint \req{p_constraint}.  Therefore the flow across the wall  from low to high mass vaccua is nearly zero and essentially all particles escape from the heavy vacuum and are trapped in the light vacuum i.e. $\rho_1\approx 1$ and $\mu_2\rightarrow-\infty$.  

The pressure on the low mass side can be approximated by the bulk pressure, as essentially all particles are elastically reflected.  On the high mass side the pressure is nearly zero since there are very few particles as compared to the low mass side.  Therefore the osmotic pressure is well approximated by the bulk pressure on the low mass side.

When $v_2$ is significantly larger than $v_1$,  the bulk pressure on the high mass side is greatly suppressed as compared to the low mass side, independently of the value of $\mu_2$ i.e. the result does not depend on sending $\mu_2\rightarrow -\infty$.  Hence the bulk pressure differential is also approximately equal to the bulk pressure on the low mass side. Therefore the osmotic effect is again negligible in this regime and the computations of the previous sections that used the bulk pressure are justified.

 Though  the osmotic pressure effect as shown here  does not  modify the  dynamics of Higgs vacuum domain walls, the computation we presented  points towards the potential for interesting extensions of the relativistic osmotic pressure model  to related areas of cosmology  and physics in general where, in the presence of multiple scales, the cancellation of membrane permeability is not expected to occur.

%%%%%%%%%%%%%%%%%%%%%%%%%%%%%
\section{Discussion and Consequences}\label{last}
This  study has shown that the capture of the Universe into a false  low mass $v_1<v_2$ ground state is a natural situation  since the high mass state $v_2$   is not the favored state at high temperature, where the Higgs vacuum state is frozen in.  Given two vacuua $v_1<v_2$, even if $v_2$ is the true vacuum at zero temperature,  $v_2$ tends to be the false vacuum for the majority of the history of the Universe. By the time $\langle h\rangle = v_H=v_1$  turns into  the false vacuum, the barrier to $\langle h\rangle = v_2$ is insurmountable, as $v_1$ emerges as `false'  only at very low -- in comparison to the EW phase transition --  temperatures. 

When does the prevailing vacuum emerge as the false vacuum? Our study shows that  the false vacuum within the `recent' past: if electrons are driving the swap, the prevailing vacuum emerges as the false vacuum   at a temperature just when BBN completes, $T_s=11$ keV.

At the end of subsection~\ref{lowT} we argued that, at temperatures that are small compared to the Higgs scale, the detailed structure of the vacuum effective potential is not important -- interpreting cosmological observations within the SM, the structure of the vacuum potential is not explored other than the location and curvature of the critical point(s), and  the vacuum energy relative to the true vacuum. 

The study we presented addresses global changes associated with the evolution of the Universe. However the appearance of two nearly degenerate minima in the potential landscape in a dynamically evolving system could on first sight also lead to incomplete swaps between the vacuum states. Here we must keep in mind that the false vacuum  is the effectively stable vacuum only due to the  presence of matter.

In subsection \ref{surfaceP} we argued that the dynamical surface pressure on the domain wall between vacuua is well approximated by the bulk pressure differential in the physically relevant regimes.  Therefore, regions of high mass vacuum would be compressed and eliminated in the Universe.  By the time the false vacuum nature of the low mass state is revealed at low temperature, one expects all remnants of the high mass vacuua to have been eliminated. This  provides a dynamical mechanism for globally freezing the Universe   into a false vacuum state characterized by low Higgs VEV, hence low particle masses,  and a nonzero dark energy density.

%%%%%%%%%%%%%%%%%%%%%%%%%%%%%
%{\bf Acknowledgments:}
\acknowledgments
This work has been supported by US Department of Energy, Office of Science, Office of Nuclear Physics  under  award number DE-FG02-04ER41318.

%\paragraph{Note added.} This is also a good position for notes added after the paper has been written.

%%%%%%%%%%%%%%%%%%%%%%%%%%%%%

\end{document}